\documentclass[preprint]{elsarticle}

\usepackage{euscript,amssymb,amsmath}

\usepackage{graphicx}
\usepackage{epsfig}
\usepackage{color}

\newcommand{\be}[1]{\begin{equation}\label{#1}}
\newcommand{\ee}{\end{equation}}
\newcommand{\ba}[1]{\begin{eqnarray}\label{#1}}
\newcommand{\ea}{\end{eqnarray}}
\newcommand{\rf}[1]{(\ref{#1})}
\newcommand{\nn}{\nonumber}

\begin{document}

\begin{frontmatter}

\title{Many-body problem in Kaluza-Klein models:\\ theory and observational consequences}

\author[a]{Alexey Chopovsky}
\ead{alexey.chopovsky@gmail.com}

\author[b,c,d]{Maxim Eingorn}
\ead{maxim.eingorn@gmail.com}

\author[b]{Alexander Zhuk}
\ead{ai.zhuk2@gmail.com}

\address[a]{Department of Theoretical Physics, Odessa National University,\\ Dvoryanskaya st. 2, Odessa 65082, Ukraine}

\address[b]{Astronomical Observatory, Odessa National University,\\ Dvoryanskaya st. 2, Odessa 65082, Ukraine}

\address[c]{Department of Theoretical and Experimental Nuclear Physics,\\ Odessa National Polytechnic University, Shevchenko av. 1, Odessa 65044, Ukraine\\}

\address[d]{North Carolina Central University,\\ Fayetteville st. 1801, Durham, North Carolina 27707, USA}

\begin{abstract}
We consider a system of gravitating bodies in Kaluza-Klein models with toroidal compactification of extra dimensions. To simulate the  astrophysical objects
(e.g., our Sun and pulsars) with the energy density much greater than the pressure, we suppose that these bodies are pressureless in the external/our space. At
the same time, they may have nonzero parameters $\omega_{(\bar\alpha -3)} \, (\bar\alpha =4,\ldots ,D) $ of the equations of state  in the extra dimensions. We
construct the Lagrange function of this many-body system for any value of $\Sigma =\sum_{\bar\alpha} \omega_{(\bar\alpha -3)}$. Moreover, the gravitational
tests (PPN parameters, perihelion/periastron advance) require negligible deviation from the latent soliton value $\Sigma =-(D-3)/2$. However, the presence of
pressure/tension in the internal space results necessarily in the smearing of the gravitating masses over the internal space and in the absence of the KK
modes. This looks very unnatural from the point of quantum physics.
\end{abstract}

\begin{keyword} extra dimensions \sep Kaluza-Klein models \sep toroidal compactification \sep tension \sep many-body problem, gravitational tests \end{keyword}


\end{frontmatter}

\section{Introduction}

The idea of multidimensionality of our Universe demanded by the theories of unification of the fundamental interactions is one of the most breathtaking ideas
of theoretical physics. It takes its origin from the pioneering papers by Th. Kaluza and O. Klein \cite{KK}, and now the most self-consistent modern theories
of unification such as superstrings, supergravity and M-theory are constructed in spacetimes with extra dimensions (see, e.g., \cite{Polchinski}). Different
aspects of the idea of multidimensionality are intensively used in numerous modern articles.

Therefore, it is important to find experimental evidence for the existence of the extra dimensions. For example, one of the aims of Large Hadronic Collider
consists in detecting of Kaluza-Klein (KK) particles which correspond to excitations of the internal spaces (see, e.g., \cite{KKparticles}).

On the other hand, if we can show that the existence of the extra dimensions is contrary to observations, then these theories are
prohibited.

In our previous papers \cite{EZ3,EZ4,EZ5} devoted to KK models with toroidal compactification of the extra dimensions, we have shown that gravitating masses
should have tension in the internal space to be in agreement with gravitational experiments in the Solar system. For example, black strings/branes with the
parameter $\omega =-1/2$ of the equation of state in the internal space satisfy this condition. For this value of $\omega$, the variations of the internal
space volume are absent \cite{EZ6}. In the dust-like case with $\omega =0$, such variations generate the fifth force, that leads to contradictions with the
experimental data.

It is worth noting that black strings/branes generalize the Schwarzschild solution to the multidimensional case. Obviously, any multidimensional theory should
have such solutions, as they must correspond to the observed astrophysical objects. Black strings/branes have toroidal compactification of the internal spaces.
This compactification is the simplest among the possible ones. However, it makes sense to investigate such models because they may help to reveal new important
properties for more physically reliable multidimensional models. The ADD model \cite{ADD1} presents a good example of it. Even if the authors use the
localization of the Standard model fields on a brane, they explore the toroidal compactification of the internal space to get the relation between the
multidimensional and four-dimensional gravitational constants \cite{ADD2}. That gives a possibility to solve the hierarchy problem and to introduce the notion
of large extra dimensions. We will not use the brane approach for our model remaining within the standard Kaluza-Klein theory. However, even in this case the
large extra dimensions can be achieved for KK models with toroidal compactification \cite{EZ2}.

In the present Letter, we try to construct the Lagrange function for a many-body system in the case of toroidal compactification. We need this function, e.g.,
to calculate the formula for advance of periastron in the case of a binary system. The measurement of this advance for the pulsar PSR B1913+16 was performed
with very high accuracy. Therefore, such measurements can be a very good test for gravitational theories. From our previous papers \cite{EZ3,EZ4,EZ5} we know
that gravitating bodies should have pressure/tension in the extra dimensions to satisfy the observable data for the deflection of light and the experimental
restrictions for the parameterized post-Newtonian parameter (PPN) $\gamma$. In this regard, the question arises about the possibility of building a many-body
Lagrange function in the presence of pressure/tension in the extra dimensions. To answer this question, we need the  metric components $g_{00}$ up to
$O(1/c^4)$, $g_{0\alpha}$ up to $O(1/c^3)$ and $g_{\alpha\beta}$ up to $O(1/c^2)$. It is worth noting that for the expressions of the deflection of light and
the PPN parameter $\gamma$, it is sufficient to calculate the metric coefficients up to $O(1/c^2)$. Obviously, the agreement with observations up to $O(1/c^2)$
does not guarantee the agreement up to $O(1/c^4)$. Hence, we calculate the metric coefficients in the required orders of $1/c$. We demonstrate that the
many-body Lagrange function can be constructed for any value of $\Sigma$ where $\Sigma$ is a sum of the parameters of the equations of state in the extra
dimensions. We show that the gravitational tests (the PPN parameter $\gamma$, and perihelion/periastron advance) allow very small deviation from the latent
soliton value $\Sigma =-(D-3)/2 \neq 0$. We prove that nonzero $\Sigma$ leads necessarily to the uniform smearing of the gravitating masses over the internal
space. However, uniformly smeared gravitating bodies cannot have  excited KK states (KK particles), which looks unnatural from the point of quantum mechanics.
In our opinion, this is a big disadvantage of the Kaluza-Klein models with the toroidal compactification.

The Letter is structured as follows. In Section 2, we provide the general description of the model and describe the nonrelativistic gravitational field for a
many-body system. The Lagrange function for this system is constructed in Section 3. The formulas for PPN parameters $\beta, \gamma$ and perihelion and
periastron advances are calculated in Section 4. The main results are briefly summarized in concluding Section 5.


\section{Nonrelativistic gravitational field for a many-body system}

To construct the Lagrange function of a system of $N$ massive bodies in $(D+1)$-dimensional spacetime, we define first the nonrelativistic gravitational field
created by this system. To do it, we need to get the metric coefficients in the weak field limit. The general form of the multidimensional metric is
\be{1}
ds^2=g_{ik}dx^idx^k=g_{00}\left(dx^0\right)^2+2g_{0\mu}dx^0dx^{\mu}+g_{\mu\nu}dx^{\mu}dx^{\nu}\, ,
\ee
where the Latin indices $i,k = 0,1,\ldots ,D$ and the Greek indices  $\mu ,\nu = 1,\ldots ,D$.
We make the natural assumption that in the case of the absence of matter sources the spacetime is Minkowski spacetime: $g_{00}=\eta_{00}=1$,
$g_{0\mu}=\eta_{0\mu}=0$, $g_{\mu\nu}=\eta_{\mu\nu}=-\delta_{\mu\nu}$. In our letter, we consider in detail the case where the extra dimensions
have the topology of tori. In the presence of matter, the metric is not Minkowskian one, and we investigate it in the weak field limit. It means that the
gravitational field is weak and velocities of test bodies are small compared with the speed of light $c$. In the weak field limit the metric is only slightly
perturbed from its flat spacetime value. We will define the metric \rf{1} up to $1/c^2$ correction terms. Because the coordinate $x^0=c t$, the metric
coefficients can be expressed as follows:
\be{2} g_{00}\approx1+h_{00}+f_{00},\quad g_{0\mu}\approx h_{0\mu}+f_{0\mu},\quad g_{\mu\nu}\approx-\delta_{\mu\nu}+h_{\mu\nu}\, , \ee
where $h_{ik}\sim O(1/c^2), f_{00}\sim O(1/c^4)$  and $f_{0\mu}\sim O(1/c^3)$. In particular, $h_{00} \equiv 2\varphi /c^2$. Later we will demonstrate that
$\varphi $ is the nonrelativistic gravitational potential.
To get these correction
terms, we should solve (in the corresponding orders of $1/c$) the multidimensional Einstein equation
\be{3}
R_{ik}=\frac{2S_D\tilde G_{\mathcal{D}}}{c^4}\left(T_{ik}-\frac{1}{D-1}g_{ik}T\right)\, ,
\ee
where $S_D=2\pi^{D/2}/\Gamma (D/2)$ is the total solid angle (the surface area of the $(D-1)$-dimensional sphere of the unit radius), $\tilde G_{\mathcal{D}}$
is the gravitational constant in the $(\mathcal{D}=D+1)$-dimensional spacetime. We consider a system of $N$ discrete massive (with rest masses $m_p,\,
p=1,\ldots ,N$) bodies. We suppose that the pressure of these bodies in the external three-dimensional space is much less than their energy density. This is a
natural approximation for ordinary astrophysical objects such as our Sun. For example, this approach works well for calculating the gravitational experiments
in the Solar system \cite{Landau}. In the case of pulsars, the pressure is not small but still much less than the energy density, and the pressureless approach
was used in General Relativity to get the formula of the periastron advance \cite{Will2}. Therefore, the gravitating bodies are pressureless in the
external/our space. On the other hand, we suppose that they may have pressure in the extra dimensions. Therefore, nonzero components of the energy-momentum
tensor of the system can be written in the following form:
\ba{4}
&{}&T^{ik}=\tilde\rho c^2u^iu^k, \quad i,k=0,\ldots ,3\, ,\\
&{}&\label{5} T^{i\bar\alpha}=\tilde\rho c^2 u^iu^{\bar\alpha}\, ,\quad
i=0,\dots ,3;\; \bar{\alpha}=4,\ldots ,D\, ,\\
&{}&\label{6} T^{\bar\alpha\bar\beta}=-p_{(\bar\alpha-3)}g^{\bar\alpha\bar\beta}+\tilde\rho c^2 u^{\bar\alpha} u^{\bar\beta}\, ,\quad
\bar{\alpha},\bar{\beta}=4,\ldots ,D\, \ea
where the $(D+1)$-velocity $u^i=dx^i/ds$ and
\be{7} \tilde\rho \equiv \sum_{p=1}^N\left[(-1)^Dg\right]^{-1/2} m_p\sqrt{g_{lm}\cfrac{dx^l}{dx^0}\frac{dx^m}{dx^0}}\delta({\bf x}-{\bf x}_p)\, , \ee
where ${\bf x}_p$ is a $D$-dimensional radius-vector of the $p$-th particle.
In what follows, the Greek indices $\alpha,\beta = 1,2,3$; $\bar\alpha,\bar\beta = 4,\ldots ,D$ and $\mu,\nu$ still run from 1 to $D$.
In the extra dimensions we suppose the equations of state:
\be{8}
p_{(\bar\alpha-3)}=\omega_{(\bar\alpha-3)}\tilde\rho c^2\, .
\ee
If all parameters $\omega_{(\bar\alpha-3)}=0$, then we come back to the model considered in our paper \cite{EZ3}. Here, massive bodies have dust-like equations
of state in all spatial dimensions. If all $\omega_{(\bar\alpha-3)}=-1/2$ (tension in the extra dimensions), then these equations of state correspond to black
strings (in the case of one extra dimension, i.e. $D=4$) and black branes (for $D>4$). If parameters satisfy the condition
$\sum\limits_{\bar\alpha}\omega_{(\bar\alpha-3)}\equiv\Sigma = - (D-3)/2$, then this case corresponds to latent solitons \cite{EZ5}. Obviously, black
strings/branes satisfy this condition.

Now, we will solve the Einstein equation \rf{3} in the same way as it was done in \cite{EZ3}. Obviously, for $\omega_{(\bar\alpha-3)}=0\, ,\bar\alpha =4,\ldots
,D$, we should reproduce the results of this paper. Up to $O(1/c^2)$, we get the following nonzero components
\ba{9}
&{}&h_{00} =\frac{2\varphi}{c^2},\quad
h_{\alpha\beta}=\frac{1-\Sigma}{D-2+\Sigma}\,\frac{2\varphi}{c^2}\delta_{\alpha\beta},\\
&{}&\label{10} h_{\bar\alpha\bar\beta}= \frac{ \omega_{(\bar\alpha-3)}(D-1) +1-\Sigma}{D-2+\Sigma}\, \frac{2\varphi}{c^2}\delta_{\bar\alpha\bar\beta}\, ,\ea
where $\varphi$ satisfies the $D$-dimension Poisson equation
\be{11}
\triangle_D\varphi({\bf x})=S_{D}\tilde G_{\mathcal{D}}\cfrac{2(D-2+\Sigma)}{D-1}\,\rho({\bf x})\, .
\ee
Here, $\triangle_D=\delta^{\mu\nu}\partial^2/\partial x^{\mu}\partial x^{\nu}$ and
the rest mass density is
\be{12}
\rho({\bf x})=\sum\limits_{p=1}^Nm_p\delta({\bf x}-{\bf x}_p)\, .
\ee
To get the solutions \rf{9} and \rf{10}, we use the standard (see, e.g., Eq. (105.10) in \cite{Landau}) gauge condition
$\partial_k\left(h^k_i-\cfrac{1}{2}\,h_l^l\delta_i^k\right)=0$. It can be easily verified that this condition is satisfied (up to $O(1/c^2)$) for $i=0,\alpha$.
However, for $i=\bar\alpha$, we should demand either $\omega_{(\bar\alpha-3)}=0$ or $\partial_{\bar \alpha} \varphi=0$. Because we consider the general case
$\omega_{(\bar\alpha-3)}\neq 0$, we must choose the latter condition. Therefore, the presence of nonzero pressure/tension in the extra dimensions results in
the metric coefficients which do not depend on the coordinates of the internal space, i.e. the gravitating masses should be uniformly smeared over the extra
dimensions. In this case, the rest mass density \rf{12} should be rewritten in the form: $\rho({\bf x})\rightarrow \rho({\bf r})=\sum_pm_p\delta({\bf r}-{\bf
r}_p)/\prod_{\bar\alpha}a_{(\bar\alpha -3)}$, where ${\bf r}_p$ is a three-dimensional radius vector of the $p$-th particle in the external space,
$a_{(\bar\alpha -3)}$ are periods of the tori (i.e. $\prod_{\bar\alpha}a_{(\bar\alpha -3)}$ is the volume of the internal space). Then, Eq. \rf{11} is reduced
to the ordinary three-dimensional Poisson equation
\be{13}
\triangle_3\varphi({\bf r})=4\pi G_N\sum_{p}m_p\delta({\bf r}-{\bf r}_p)
\ee
with the solution
\be{14}
\varphi({\bf r})=-\sum_{p}\frac{G_N m_p}{|{\bf r}-{\bf r}_p|}\, ,
\ee
where $G_N$ is the Newtonian gravitational constant:
\be{15}
4\pi G_N =\frac{2S_D(D-2+\Sigma)}{(D-1)\prod_{\bar\alpha}a_{(\bar\alpha -3)}} \tilde G_{\mathcal{D}}\, .
\ee

Following the paper \cite{EZ3}, we can also obtain the $O(1/c^3)$ and $O(1/c^4)$ correction terms. As a result, the metric coefficients read
\ba{16}
g_{00}&\approx& 1+\frac{2\varphi({\bf r})}{c^2}+\frac{2\varphi^2({\bf r})}{c^4}+\frac{2G_N^2}{c^4}
\sum_p\cfrac{m_p}{|{\bf r}-{\bf r}_p|}\sum_{q\neq p}\frac{m_q}{|{\bf r}_p-{\bf
r}_q|}\nn\\
&-&\frac{D-\Sigma}{D-2+\Sigma}\frac{G_N}{c^4}\sum_p\frac{m_pv_p^2}{|{\bf r}-{\bf r}_p|}\, , \ea
\be{17} g_{0\alpha}\approx\frac{3D-2-\Sigma}{D-2+\Sigma}\frac{G_N}{2c^3}\sum_p\frac{m_p}{|{\bf r}-{\bf
r}_p|}\,v^{\alpha}_p+\frac{G_N}{2c^3}\sum_p\frac{m_p}{|{\bf r}-{\bf r}_p|}\, n_p^{\alpha} ({\bf n}_p{\bf v}_p)\, , \ee
\be{18}
g_{\alpha\beta}\approx\left(-1+\cfrac{1-\Sigma}{D-2+\Sigma}\cfrac{2\varphi({\bf r})}{c^2}\right) \delta_{\alpha\beta}\, ,
\ee
\be{19}
g_{\bar\alpha\bar\beta}\approx\left(-1+\cfrac{\omega_{(\bar\alpha-3)}(D-1)+1-
\Sigma}{D-2+\Sigma}\, \cfrac{2\varphi({\bf r})}{c^2}\right)\delta_{\bar\alpha\bar\beta}\, .
\ee
Here, the three-velocity $v^{\alpha}_p=dx^{\alpha}_p/dt$, the three-dimensional unit vector $n_p^{\alpha}=(x^{\alpha}-x_p^{\alpha})/|{\bf r}-{\bf r}_p|$ and
$({\bf n}_p{\bf v}_p)=\sum_{\alpha}n_p^{\alpha}v^{\alpha}_p$. Obviously, if $\omega_{(\bar\alpha-3)}=0,\; \forall \; \bar\alpha,\ \Rightarrow\ \Sigma =0$,
these formulas are reduced to ones in \cite{EZ3}, which in turn coincide with the metric coefficients obtained in \cite{Landau} in the case $D=3$.


\vspace{0.3cm}

\section{Lagrange function for a two-body system}

Let us construct now the Lagrange function of the many-body system described above. To perform it we will follow the procedure described in \cite{Landau} (see
\S 106). The Lagrange function of a particle $p$ with the mass $m_p$ in the gravitational field created by the other bodies is given by the expression
\be{20} L_p=-m_pc\cfrac{ds_p}{dt}=-m_p c^2\left(g_{00}+2\sum_\mu g_{0\mu}\cfrac{v^\mu_p}{c}+\sum_{\mu\nu}g_{\mu\nu}\cfrac{v^\mu_pv^\nu_p}{c^2}\right)^{1/2}\, ,
\ee
where the metric coefficients are taken at ${\bf r}={\bf r}_p$. For our purposes, it is sufficient to consider the case of two particles.
The substitution of the metric coefficients \rf{16}-\rf{19} leads to the following expression for the particle "1":
\ba{21}
L_1&=&\mathrm{f}({\bf v}_1^2)+G_N\cfrac{m_1m_2}{|{\bf r}-{\bf r}_2|}-\cfrac{1}{2c^2}\,G_N^2\cfrac{m_1m_2^2}{|{\bf r}-{\bf r}_2|^2}
-\frac{1}{c^2}G_N^2
\cfrac{m_1^2m_2}{|{\bf r}-{\bf r}_2||{\bf r}_1-{\bf r}_2|}\nn\\
&+&\cfrac{1}{2c^2}\cfrac{G_Nm_1m_2}{|{\bf r}-{\bf
r}_2|}\left[\mathfrak{a}(D,\Sigma)v_2^2+2\left(\mathfrak{b}(D,\Sigma)+\cfrac{1}{2}\right)v_1^2\right.\nn\\
&-&\left.\mathfrak{c}(D,\Sigma)({\bf v}_1{\bf v}_2)-({\bf n}_2{\bf v}_2)({\bf
n}_2{\bf v}_1)\right]\, ,
\ea
where $\mathrm{f}({\bf v}_1^2)=m_1v_1^2/2+m_1v_1^4/(8c^2)$ and we drop the term $-m_1 c^2$. Here, we use the following abbreviations:
\ba{22}
\mathfrak{a}(D, \Sigma)&\equiv&\cfrac{D-\Sigma}{D-2+\Sigma}, \quad \mathfrak{b}(D, \Sigma)\equiv\cfrac{1-\Sigma}{D-2+\Sigma}\, ,\nn\\
\mathfrak{c}(D, \Sigma)&\equiv&\cfrac{3D-2-\Sigma}{D-2+\Sigma}\, .
\ea
The total Lagrange function of the two-body system should be constructed so that it leads to the correct values of the forces $\left.\partial L_p/\partial {\bf
r}\right|_{{\bf r}={\bf r}_p}$ acting on each of the bodies for given motion of the others \cite{Landau}. Following this prescription, we obtain from \rf{21} the
two-body Lagrange function
\ba{23}
L^{(2)}_{1}&=&\mathrm{\tilde f}({\bf v}_1^2, {\bf v}_2^2)+\cfrac{G_Nm_1m_2}{r_{12}}-\cfrac{G_N^2m_1m_2(m_1+m_2)}{2c^2r_{12}^2}\nn\\
&+&\cfrac{G_Nm_1m_2}{2c^2r_{12}}\left[\mathfrak{a}(D,\Sigma)v_2^2+\left(2\mathfrak{b}(D,\Sigma)+1\right)v_1^2\right.\nn\\
&-&\left.\mathfrak{c}(D,\Sigma)({\bf v}_1{\bf v}_2)-({\bf n}_{12}{\bf
v}_1)({\bf n}_{12}{\bf v}_2)\right]\, ,
\ea
where $\mathrm{\tilde f}({\bf v}_1^2, {\bf v}_2^2)=\sum_{a=1}^2m_av_a^2/2+\sum_{a=1}^2m_av_a^4/(8c^2)$. It can be easily seen that $\left.\partial L_1/\partial
{\bf r}\right|_{{\bf r}={\bf r}_1} = \partial L^{(2)}_1/\partial {\bf r}_1$. By the same way we can construct the two-body Lagrange function $L^{(2)}_2$ from
the Lagrange function $L_2$ for the particle "2":
\ba{24}
L^{(2)}_{2}&=&\mathrm{\tilde f}({\bf v}_1^2, {\bf
v}_2^2)+\cfrac{G_Nm_1m_2}{r_{12}}-\cfrac{G_N^2m_1m_2(m_1+m_2)}{2c^2r_{12}^2}\nn\\
&+&\cfrac{G_Nm_1m_2}{2c^2r_{12}}\left[\mathfrak{a}(D,\Sigma)v_1^2+\left(2\mathfrak{b}(D,\Sigma)+1\right)v_2^2\right.\nn\\
&-&\left.\mathfrak{c}(D,\Sigma)({\bf v}_1{\bf v}_2)-({\bf n}_{12}{\bf
v}_1)({\bf n}_{12}{\bf v}_2)\right]\, .
\ea
It is worth noting that both $L^{(2)}_{1}$ and $L^{(2)}_{2}$ are reduced to the Lagrange function of the two-body system in \cite{Landau} if we assume that
$D=3$, $\Sigma=0$.

Obviously, the Lagrange functions $L^{(2)}_{1}$ and $L^{(2)}_{2}$ should be symmetric with respect to permutations of particles 1 and 2 and should coincide
with each other. This requires the following condition:
\be{25}
\mathfrak{a}(D,\Sigma)=2\mathfrak{b}(D,\Sigma)+1\, ,
\ee
which is satisfied identically for any value of $\Sigma$. Therefore, we construct the two-body Lagrange function for any value of the parameters of the
equations of state in the extra dimensions.


\vspace{0.3cm}

\section{Gravitational tests}

It can be easily seen that the metric coefficients in the external/our space \rf{16}-\rf{18} as well as the two-body Lagrange functions \rf{23} and \rf{24}
exactly coincide with the corresponding expressions in General Relativity for the value $\Sigma = \sum_{\bar\alpha}\omega_{(\bar\alpha-3)} = - (D-3)/2$, i.e.
in the latent soliton case \cite{EZ5}. Black strings/branes are particular cases of this class. Therefore, the known gravitational tests in this case give the
same results as for General Relativity. In other words, we get good agreement with observations. It is of interest to obtain an experimental restriction on a
deviation from this value. For this purpose, we write $\Sigma$ in the following form:
\be{26}
\Sigma = -\frac{D-3}{2}+\varepsilon
\ee
and find the experimental limitations on $\varepsilon$.

\

{\it PPN parameters}

\vspace{0.2cm}

To get the parameterized post-Newtonian (PPN) parameters $\beta$ and $\gamma$, we consider the case of one particle at rest. Then, we can easily obtain from
Eqs. \rf{16} and \rf{18} that
\be{27}
\beta =1\, ,\quad \gamma = \frac{1-\Sigma}{D-2+\Sigma}\, ,
\ee
i.e. the PPN parameter $\beta$ exactly coincides with the value in the General Relativity. There are strong experimental restrictions on the value of $\gamma$.
The tightest constraint on $\gamma$ comes from  the Shapiro time-delay experiment using the Cassini spacecraft, namely: $\gamma-1 =(2.1\pm 2.3)\times 10^{-5}$
\cite{Will2,JKh,Bertotti}. In our case
\be{28}
\gamma - 1 \approx -\frac{4\varepsilon}{D-1}\, .
\ee
Therefore, the Shapiro time-delay experiment results in the following limitation:
\be{29} |\varepsilon| \lesssim (D-1)\times 10^{-5}\, . \ee

\

{\it Perihelion shift of the Mercury}

\vspace{0.2cm}

For a test body orbiting around the gravitating mass $m$, the perihelion shift for one period is given by the formula \cite{Will2,Will}
\be{30}
\delta \psi = \frac13 \left(2+2\gamma -\beta\right)\, \frac{6\pi G_N m}{c^2a(1-e^2)}\equiv
\frac13 \left(2+2\gamma -\beta\right)\, \delta \psi_{GR}\, ,
\ee
with $a$ and $e$ being the semi-major axis and the eccentricity of the ellipse, respectively. $\delta \psi_{GR}$ is the value for General Relativity. In the
case of Mercury this calculated value is equal to 42.98 arcsec per century \cite{Will2,Will3}. This predicted relativistic advance agrees with the observations
to about 0.1\% \cite{Will2}. Substituting the PPN parameters \rf{27} in this formula, we obtain the advance in our case:
\be{31}
\delta \psi = \frac13 \frac{D-\Sigma}{D-2+\Sigma}\, \delta \psi_{GR} \approx \left(1-\frac{8}{3(D-1)}\varepsilon\right)\delta \psi_{GR}\, .
\ee
Obviously, to be in agreement with the observation not worse than General Relativity, the parameter $\varepsilon$ should satisfy the condition
\be{32}
|\varepsilon| \lesssim \frac{3(D-1)}{8}\times 10^{-3}\, .
\ee
Therefore, this limitation is less strong than \rf{29}.


\

{\it Periastron shift of the relativistic binary pulsar
PSR B1913+16}

\vspace{0.2cm}

Much more strong limitation can be found from the measurement of the periastron shift of the relativistic binary pulsar. First, the advance of periastron in
these systems is in many orders of magnitude bigger than for the Mercury. Second, the measurements are extremely accurate. For example, for the pulsar PSR
B1913+16 the shift is $4.226598\pm 0.000005$ degree per year \cite{pulsar}. For such system both the pulsar and the companion have comparable masses. In the
case of General Relativity, a solution for orbital parameters yields mass estimates for the pulsar and its companion, $m_1 = 1.4398\pm 0.0002 M_{\odot}$ and
$m_2 = 1.3886 \pm 0.0002 M_{\odot}$, respectively. It is worth noting that these are calculated values (not observable!) which are valid for General
Relativity. Because two bodies have comparable masses, to get a formula for the advance we need a two-body Lagrangian. Then, following the problem 3 in \S 106
\cite{Landau} we get for our two-body Lagrangians \rf{23} and \rf{24} the desired formula in the form of \rf{31} with the well known General Relativity
expression
\be{33}
\delta \psi_{GR} = \frac{6\pi G_N (m_1+m_2)}{c^2a(1-e^2)}\, .
\ee
In future, independent measurements of masses $m_1$ and $m_2$ will allow us to obtain a high accuracy restriction on the parameter $\varepsilon$.


\vspace{0.3cm}

\section{Summary}

In this Letter, we have constructed the Lagrange function for a two-body system in the case of Kaluza-Klein models with toroidal compactification of the extra
dimensions. The case of more than two bodies is straightforward. We supposed that gravitating bodies are pressureless in the external/our space. This is a
natural approximation for ordinary astrophysical objects such as our Sun. For example, this approach works well for calculating the gravitational experiments
in the Solar system \cite{Landau}. In the case of pulsars, the pressure is not small but still much less than the energy density. Hence, the pressureless
approach is used in General Relativity to get the formula \rf{33} which is in very good agreement with the observations of advance of periastron of the pulsar
PSR B1913+16.

With respect to the internal space, we supposed that  gravitating masses may have nonzero parameters $\omega_{(\bar\alpha -3)} \, (\bar\alpha =4,\ldots ,D) $
of the equations of state  in the extra dimensions. We have shown that the Lagrange function of this many-body system can be constructed for any value of the
parameter $\Sigma =\sum_{\bar\alpha} \omega_{(\bar\alpha -3)}$.

To construct the many-body Lagrangian, as well as to get the formulas for the gravitational tests, we obtained the  metric components $g_{00}$ up to
$O(1/c^4)$, $g_{0\alpha}$ up to $O(1/c^3)$ and $g_{\alpha\beta}$ up to $O(1/c^2)$. These expressions exactly coincide with the corresponding formulas in
General Relativity for the value $\Sigma = \sum_{\bar\alpha}\omega_{(\bar\alpha-3)} = - (D-3)/2$. This is the latent soliton case \cite{EZ5}. Black
strings/branes are particular cases of it with all $\omega_{(\bar\alpha -3)}=-1/2\,\ \ \forall \bar\alpha$. Obviously, the known gravitational tests (PPN
parameters, perihelion/periastron shift) in this case give the same results as for General Relativity. On the other hand, we used these tests to get the
restrictions on the deviation from the latent soliton value. At the present, the most strong restriction follows from the time delay of radar echoes (the
Cassini spacecraft mission). The two-body Lagrange function allowed us to get the formula for the advance of the periastron. In future, when the masses of the
binary pulsar system PSR B1913+16 will be measured, the advance of this periastron can be used to get the restriction with very high accuracy. All obtained
limitations indicate very small deviation from the latent soliton value. Therefore, the pressureless case $\Sigma =0$ in the internal space is forbidden, in
full agreement with the results of the paper \cite{EZ3}.  This conclusion does not depend on the size of extra dimensions. The physical reason of it is that in
the case of toroidal compactification, only in the case of latent solitons the variations of the total volume of the internal space are absent \cite{EZ6}.

One more important result obtained in this latter is worth noting. As we have shown above (see also \cite{EZ4,EZ5,EZ6}), tension in the internal spaces is the
necessary condition to satisfy the gravitational experiments in KK models with toroidal compactification. In our letter, we have proven that the presence of
pressure/tension in the internal space leads necessarily to the uniform smearing of the gravitating masses over the internal space. For example, black
strings/branes have tension in the internal space (see, e.g., \cite{Traschen}). Therefore, they should be smeared. However, uniformly smeared gravitating
bodies cannot have  excited KK states (KK particles), which looks unnatural from the point of quantum mechanics. In our opinion, this is a big disadvantage of
the Kaluza-Klein models with the toroidal compactification. It is of interest to check this property for models with other types of compactification (e.g.
Ricci-flat, spherical). This is the subject of our subsequent study.



\section*{Acknowledgements}

This work was supported in part by the "Cosmomicrophysics-2" programme of the Physics and Astronomy Division of the National Academy of Sciences of Ukraine.
The work of M. Eingorn was supported by NSF CREST award HRD-0833184 and NASA grant NNX09AV07A.




\end{document}